\documentclass[runningheads,envcountsame,a4paper]{llncs}

\usepackage[T1]{fontenc}
\usepackage[latin2]{inputenc}
\usepackage{amsmath}
\usepackage{latexsym}

\usepackage{graphicx}

\usepackage{color}

\def\dt{{\rm d}\,}

\newcommand{\ket}[1]{| #1 \rangle}
\newcommand{\bra}[1]{\langle #1 |}

\renewcommand{\rho}{\varrho}

\def\duzomniejsze{<\kern-.7mm<}
\def\duzowieksze{>\kern-.7mm>}

\def\beq{\begin{equation}}
\def\eeq{\end{equation}}
\def\be{\begin{equation}}
\def\ee{\end{equation}}
\def\ben{\begin{eqnarray}}
\def\een{\end{eqnarray}}
\def\beqa{\begin{eqnarray}}
\def\eeqa{\end{eqnarray}}
\def\eea{\end{array}}
\def\bea{\begin{array}}
\newcommand{\bei}{\begin{itemize}}
\newcommand{\eei}{\end{itemize}}
\newcommand{\bee}{\begin{enumerate}}
\newcommand{\eee}{\end{enumerate}}

\def\1{\openone}

\def\>{\rangle}
\def\<{\langle}

\def\dt#1{{{\kern -.0mm\rm d}}#1\,}
\def\squareforqed{\hbox{\rlap{$\sqcap$}$\sqcup$}}
\def\qed{\ifmmode\squareforqed\else{\unskip\nobreak\hfil
\penalty50\hskip1em\null\nobreak\hfil\squareforqed
\parfillskip=0pt\finalhyphendemerits=0\endgraf}\fi}

\def\bep{\begin{proposition}}
\def\eep{\end{proposition}}
\def\bel{\begin{lemma}}
\def\eel{\end{lemma}}

\def\bet{\begin{theorem}}
\def\eet{\end{theorem}}
\def\bed{\begin{definition}}
\def\eed{\end{definition}}
\def\bef{\begin{fact}}
\def\eef{\end{fact}}

\begin{document}
\title{The fragility of quantum information?}
\author{Barbara M. Terhal}
\institute{Institute for Quantum Information, RWTH Aachen University, 52056 Aachen, Germany,
\email{terhal@physik.rwth-aachen.de}}

\maketitle

\begin{abstract}
We address the question whether there is a fundamental reason why quantum information is more fragile than classical information. We show that 
some answers can be found by considering the existence of quantum memories and their dimensional dependence.  
\end{abstract}

\section{Classical information}

To store a bit of information robustly, redundancy is needed. The bits of information recorded in Egyptian hieroglyphs, written on the walls of Egyptian temples, have been preserved for over 5000 years. 
% Karnak temple
The characters can still be 
recognized and deciphered as they are captured by a macroscopic displacement of hard stone, a carving. Microscopic details, the particular placement of the sandstone particles, the particular quantum states of the ${\rm  SiO_2}$ molecules
and their binding together in a crystal, have been thoroughly altered by the weather over centuries. However the microscopic details do not affect the macroscopic message as long as the variations of the microscopic details are small, random and do not accumulate over time 
to lead to macroscopic changes. In this sense, the encoded information is a phenomenon which robustly {\em emerges} from an underlying statistical reality \cite{book:anderson}. And when we decypher such glyphs, we error-correct: a carving of an owl with an obliterated head is still sufficiently different from a feather hieropglyph, as what makes the owl an owl is redundantly present in the hieroglyph. What this one example teaches us, is that (1) classical information can be stored for incredibly long times in the presence of a steadily-acting physical environment, that (2) we need to error-correct upon read-out, i.e. ignore the small fluctuations and (3) there will always be events which can destroy the information, easily and decisively, e.g. hacking into the stone by iconoclasts or the demolition of the entire temple, but these events can be assumed to be rare and non-random.

% O(10^-10) is length of unit cell in atomic lattice 
% ferrite magnetite? ferrimagnetic 

Of course, this principle of robust storage still underlies our modern computer technology, for example in the form of the hard-disk drive. The jump from hieroglyphs to hard-disk drives is a huge jump in storage density, 
% 1000Gbit/inch, 6,4 x 10^-4 m^2 for 10^12 bits, so 6.4 x 10^{-16} per bit
from, say, $5$ bits per square inch to almost $10^{12}$ bits per square inch. This jump is only possible, because even in an area of $10^{-8} \times 10^{-8}\,{\rm m^2}$, a macrosopic number of degrees of freedom, namely electrons and their spins, are swimming around to provide redundancy. The encoding of a single bit "0" or "1" is done using tens of magnetic grains in a ferromagnetic material (for example, gamma ferric oxide ${\rm \gamma\mbox{-}Fe_2O_3}$). The bit "1" is represented by a domain wall in the ensemble of grains at which the sign of the magnetization changes whereas the "0" bit is repesented by a uniform magnetization of the underlying grains. The stability of the magnetization of a single grain is due to the phenomenon of ferromagnetism: even in the absence of a magnetic field the electron spins and their angular momenta in the crystalline atomic structure give rise to a nonzero magnetization. Ferromagnetism is temperature-dependent, the capability to spontaneously magnetize is lost above the Curie temperature $T_c$ ($T_c=600^{\circ}$C for ${\rm \gamma\mbox{-}Fe_2O_3}$). For small magnetic grains the stability also depends on the size of the grain as the {\em energy barrier} to switch the overall magnetization depends on the volume of the grain. Smaller magnetic grains thus have an increased susceptibility to thermal fluctuations, --an effect which is called superparamagnetism--, which sets limits to miniaturization of hard-drive technology.

We see that the root cause of robustness is redundancy at the physical level of many electrons mutually aligning their magnetic moments  in magnetic domains. Why do the electrons do this? This turns out to be 
largely due to a quantum mechanical effect. Electrons in unfilled iron shells interact via the ferromagnetic Heisenberg exchange interaction, i.e. the interaction between two spins $i$ and $j$ is $- J (X_i X_j+Y_i Y_j+Z_i Z_j)$ where $X,Y,Z$ are the three Pauli matrices \footnote{$X=\left(\begin{array}{cc} 0 & 1 \\ 1 & 0 \end{array}\right), Y=\left(\begin{array}{cc} 0 & -i \\ i & 0 \end{array}\right), Z=\left(\begin{array}{cc} 1 & 0 \\ 0 & -1 \end{array}\right)$.} and $J$ is a coupling constant larger than zero. A simple model which could allow us to understand the origin of stability is that of a lattice of spin-1/2 particles, qubits, locally interacting with their neighbors by this Heisenberg exchange interaction. If the geometry of the lattice is two-dimensional, the Mermin-Wagner theorem states that there can be no spontaneous symmetry breaking, --no low-temperature phase characterized by a nonzero magnetization is possible--. This is in contrast with the situation for a three-dimensional lattice where 
spontaneous magnetization does occur for temperatures below the critical temperature. Even though the magnetic recording film is thin, it is still many atoms thick and the three-dimensional picture applies. Naturally, --as a curious child which keeps on asking--, we are led to ask why the Mermin-Wagner theorem holds in one and two dimensions and not in three dimensions. In all dimensions, at any temperature, excitations above the ferromagnetic ground-state in the form of spin waves contribute negatively to the magnetization. Whether they overwhelm the ferromagnetic order in the ground-state depends on the energy spectrum of these excitations and the number of excitations at any given energy (i.e. the free energy) which in turn depends on the dimension of the system \cite{book:am}.

That the interplay between entropy and energy cost of the excitations can be dimensionally-dependent was first understood for an even simpler model of ferromagnetism, namely the Ising model, by Peierls \cite{peierls}. In the one-dimensional (1D) Ising model the Hamiltonian on $n$ qubits is $H=- J\sum_{i=1}^{n-1} Z_i Z_{i+1}$ so that the two degenerate ground-states are $\ket{00\ldots 0}$ and $\ket{11 \ldots 1}$. A single bit can be redundantly encoded in these two ground-states.  The encoding of such bit is robust if thermal excitations do not wipe out the average magnetization $M=\frac{1}{n} \sum_i \langle Z_i \rangle$, in other words, if the system would exhibit spontaneous magnetization at non-zero temperature. In 1D this is not the case and a quick way of understanding this is to realize that there are no energy barriers for creating large domains of spins with opposite magnetization. In the language of errors and coding, one can say that there are no energy mechanisms which prevent errors from accumulating and thus error accumulation eventually leads to a reversal of the magnetization which represents a {\em logical} error on the encoded bit. We error-correct upon read-out as we consider, not the individual expectation of each operator $Z_i$, but the average magnetization $M$ which, when $M < 0$ is interpreted as signaling the bit "1", and when $M > 0$ the bit "0". The two-dimensional (2D) version of the Ising model has a different phase-diagram with a critical temperature $T_c$ separating a ferromagnetic phase where robust storage of a bit is possible from a higher-temperature paramagnetic phase. The reason for the discrepancy is dimensional. In any dimensions an excitation consisting of a domain of spins of opposite magnetization costs an energy proportional to its boundary. In two dimensions this boundary grows as the domain grows, providing a mechanism to energetically suppress the growth of excitations. 
The energy barrier, defined as the minimum energy cost of a spin configuration through which one necessarily has to pass to get from $\ket{00 \ldots 0}$ to $\ket{11 \ldots 1}$ (or vice versa), is $L$ for a two-dimensional $L \times L$ lattice. For a 1D chain of spins, the energy of an excited domain is $O(1)$ as the boundary is 0-dimensional, and thus the energy barrier is $O(1)$ for a one-dimensional chain of spins.  For the 2D Ising model the energy barrier grows with system-size, providing more robustness the larger the system. Studies of mixing times of Markov chains mimicking the interaction with a thermal environment have confirmed this basic picture, see e.g. \cite{book:mix}.

Now that we have perhaps sketched the ideas underlying the robust storage of classical information, is it time to turn to quantum bits \cite{book:nielsen&chuang}. Wouldn't it be nice to preserve a quantum bit for 5000 years? Is there a fundamental principle at play that prevents us from doing this?

\section{Quantum information, anyone?}

% bacon reference
%  \footnote{Note the difference between the ferromagnetic Heisenberg model where the spontaneous symmetry breaking concerns a 
% local continuous symmetry, namely the commutation of the Hamiltonian with $U \otimes^n$ for any $U$, and the Ising model where the symmetry, i.e. the commutation of $H$ with $X^{\otimes n}$ is discrete. In the toric model, the symmetry is continuous but non-local, i.e. $H$ commutes with the logical operators $\overline{U}=e^{i \theta \hat{n} \cdot \overline{S}}$ with $\overline{S}=(\overline{X},\overline{Y},\overline{Z})$.
% SC to do...
% low T superconductivity....
% shape anisotropy to break continuous symmetry

At first sight the idea to encode information in the states of a single atoms, electrons or a single bosonic mode, is terribly, incredibly, naive. In hindsight it could only have come about in a time as prosperous and full of hubris as the past 30 years and by people who were unburdened by their knowledge of physics. By and large in order to understand macroscopic phenomena, whether something conducts electricity or heat, whether it shines or is dull, whether it is a magnet or not, whether it superconducts, we have recourse to the quantum mechanical properties of electrons and atoms in materials. We use quantum theory to obtain a description of the emergent phenomena which are intrinsically classical; but emergent phenomena which are quantum in themselves, this is something that we may even have a hard time conceptualizing.  As we argued above, robust are those phenomena which emerge from such statistical microscopic reality due to their redundancy. States of single electron spins decohere in microseconds, electronic states in trapped-ion or nuclear spin qubits may survive for seconds. Without any back-up redundancy these qubits are bound to such smallish coherence times, whereas with redundancy, i.e. statistical ensembles, lattices, arrays of such coupled qubits we expect to be back in the world of classical phenomena. But perhaps not quite. 

The best demonstration of an emergent quantum phenonemon is superconductivity: the complex order parameter $\psi$ of the condensate is a macrosopic degree of freedom emerging from the interactions of many electrons. A superconducting flux or persistent current qubit in the state $\ket{0}$ is realized as superconducting loop with current going clock-wise, while current going counter clock-wise can represent the orthogonal state $\ket{1}$. The magnetic flux in the loop is quantized in units of $ \frac{h c}{2 e}$. Upon application of half of such unit of flux, it is energetically most favorable for the loop to carry currents, clockwise or, energetically-equivalent, counterclockwise, so that these currents make the magnetic flux an integer flux unit.  The transition from $\ket{0}$ to $\ket{1}$ can be prohibited as it would require a macrosopic change of the state of the condensate through processes which are energetically unfavorable.

This flux qubit could be operated as a classical bit in which only the states $\ket{0}$ or $\ket{1}$ are used, or as a qubit which we wish to keep in a coherent superposition $\alpha \ket{0}+\beta \ket{1}$ \cite{mooij+:flux}. In its first incarnation it has been proposed as the developing RSFQ (rapid-single-flux-quantum) technology \cite{lik_sem:rsfq} and such flux quantum bit can be preserved for, in all likelihood, many years.  If we wish to use it as a qubit then more protection from noise is required, as qubits can dephase while classical bits cannot. If a qubit is defined as two degenerate (or nondegenerate) eigenstates of a Hamiltonian, dephasing occurs whenever the energies of these two eigenstates randomly fluctuate in time. The eigenenergies are susceptible typically to any coupling of the selected two-level system with the environment: charges, spins, phonons, stray magnetic or electric fields,  even though these couplings may be relatively weak. 

One may thus be led to ask:  are there systems in which additional weak terms in the Hamiltonian do not affect the eigenenergies of, say, a degenerate ground-space in which one stores a qubit? It was the important insight of Alexei Kitaev to address this question and relate it to the concept of topology and topological order \cite{kitaev:anyons}. Kitaev envisioned using a two-dimensional, topologically-ordered material supporting anyonic excitations. By braiding of these excitations (i.e. moving them around) one would be able to perform universal logic on the qubits encoded in a degenerate eigenspace of the Hamiltonian of the material. The quest to realize this topological quantum computation scheme in a fractional quantum Hall system is ongoing.

\begin{figure}[htb]
\centerline{
\mbox{
\includegraphics[width=0.9\linewidth]{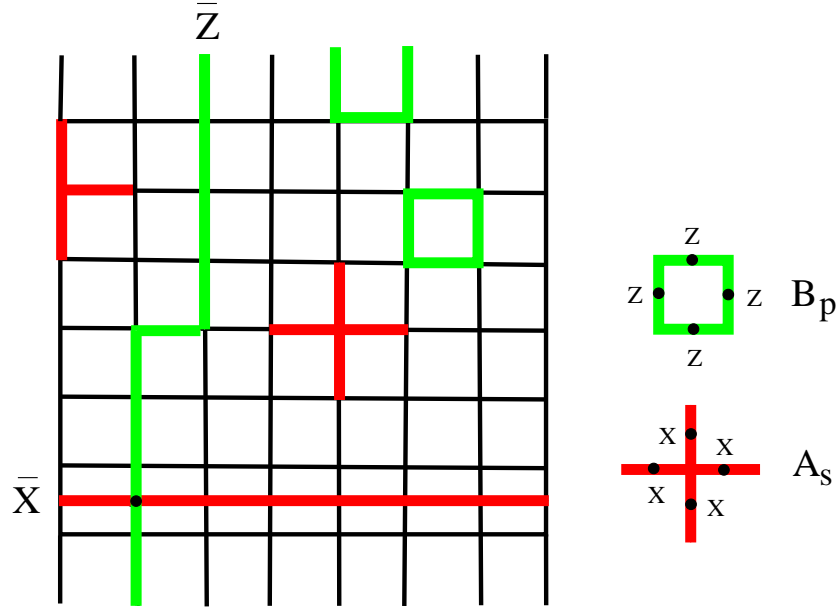}}}
\caption{Surface Code Model on a $L \times L$ lattice. A qubit lives on every {\em edge} of the black lattice, in total there are $L^2+(L-1)^2$ qubits. Two types of local parity check operators, $A_s$ and $B_p$, each act on four qubits (except at the boundary). The subspace of states which satisfy the parity checks, i.e. for which $A_s=+1$ and $B_p=+1$, is two-dimensional and hence represents a qubit. The logical $\overline{Z}$ and $\overline{X}$ operator of this qubit is any string of $Z$ operators connecting the top- to bottom (realizing $\overline{Z}$) or any string of $X$ operators connecting the left- to right boundary of the lattice (realizing $\overline{X}$).}
\label{fig:surface}
\end{figure}

Kitaev also introduces a toy model Hamiltonian related to a quantum error-correcting code, the toric code \cite{kitaev:anyons}, which although extremely simple, captures the essential features of  how may be able to store a qubit robustly in a 2D macroscopic system. The features of this model are most easily explained by considering the variant of this model which is called the surface code. We imagine elementary qubits laid out on the edges of a 2D lattice, see Fig. \ref{fig:surface}. The surface code Hamiltonian reads $H=-\Delta (\sum_s A_s+\sum_p B_p)$ where $\sum_p$ is the sum over all plaquettes on the lattice and $\sum_s$ is the sum over stars and $\Delta > 0$. The plaquette term $B_p$ equals the product of four Pauli $Z$ operators around a plaquette on the lattice. One can view this operator as a parity check on the qubits, $B_p=+1$ for even parity and $B_p=-1$ for odd parity. The star operator $A_s$ is similarly a 4-qubit parity check, but on the dual lattice, i.e. it applies to the 4 qubits on edges emanating from a vertex, and the check occurs in the Hadamard-rotated basis so that it is a product of Pauli $X$s (the Hadamard transformation $H: X \leftrightarrow Z$). At the boundaries the parity checks become 3-qubit checks, see the Figure. This Hamiltonian, like the 2D Ising model, has a two-dimensional degenerate ground-space. One can take two orthogonal states in this space and call them the logical "0" denoted as $\ket{\overline{0}}$ and "1" denoted as $\ket{\overline{1}}$.  

Can we understand how one can store a qubit as $\alpha \ket{\overline{0}}+\beta \ket{\overline{1}}$ robustly? The qubit is characterized by the logical 
operations $\overline{X}\;: \ket{\overline{0}} \leftrightarrow \ket{\overline{1}}$ and $\overline{Z}\;:\ket{\overline{0}}+\ket{\overline{1}} \leftrightarrow \ket{\overline{0}}-\ket{\overline{1}}$. The logical operator $\overline{Z}$ can be chosen to be a product of Pauli $Z$s connecting the top boundary to the bottom boundary, and the logical operator $\overline{X}$ is any path of $X$s connecting the left to the right boundary, see Figure \ref{fig:surface}. Both these operators are topological, i.e. if we impose periodic boundary conditions on the lattice, they would represent non-contractible loops, which also means that they can be freely deformed as long as they go from boundary to boundary. 

When we compare this simple model to the 2D Ising model we see two differences. In the 2D Ising model, the logical $\overline{X}$ consists of Pauli $X$s applied to the (qu)bits of the entire surface, while the logical $\overline{Z}$ is a single Pauli $Z$ on any qubit on the lattice. In the surface code, {\em both} logical operators are non-local, i.e. growth in length with the linear dimension of the lattice $L$. This difference has good and bad consequences. One can rigorously prove (see e.~g.~\cite{BHM:top_order}) that for models such as the surface code, the ground-states of $H+\epsilon V$ where $V$ is a sum of local arbitrary perturbations on the qubits on the lattice, are degenerate to exponential precision $O(\exp(-c L^{\alpha}))$ ($\alpha > 0$) as long as $\epsilon$, the strength of the perturbation, is below some critical value $\epsilon_c$. Hence the bigger the lattice, the more protection the model offers against dephasing of the encoded qubit. This is the essence of topological order: the measurement of an observable which is restricted to a subregion of the lattice will not reveal any information about the value of $\alpha$ and $\beta$ of the underlying quantum state. This information is conveniently encoded in the non-local degrees of freedom, hidden away from local disruptive processes. This notion of topological order is also the defining property of the surface code as a quantum error correcting code for which the code distance, defined as the minimum weight of any logical operator, is $L$. 

But there is bad news too, which has a dimensional dependence. The two logical operators are both strings and not surfaces as in the 2D Ising model and this directly impacts the protection of our qubit against thermal excitations, as was first noted in \cite{dennis+:top}. A thermal excitation or an error corresponds to a change in the parity check values: for a $X$ error on a certain edge, the plaquette operators $B_p$ next to this edge will flip sign from $+1$ to $-1$. A string of $X$ errors similarly terminates at two sites where the plaquette operator has flipped signs; these can be called defects or particles which can diffuse (the string of $X$ errors gets longer) or be annihilated (the $X$ error disappears). As in the 1D Ising model, a long connected string of $X$ errors has the same energy as a short string, because the price for the excitation is paid at the 0-dimensional boundary. Phrased differently, the energy barrier for thermal excitations that destroy the stored quantum information, is a constant, independent of lattice size $L$. This finding has been corroborated in \cite{AFH:toric} by showing that the storage time of such a qubit at finite temperature $T$ is $O(e^{-c \Delta/T})$ (with some constant $c$), independent of system size. This means that the possibility of robustly storing a qubit is restricted to the realm of sufficiently low temperatures $T \ll \Delta$. 

It is natural to ask what can then be achieved in three or higher dimensions. It was Franz Wegner in 1971 who considered this question in the setting of $Z_2$-gauge theories \cite{wegner:71,kogut:rmp}. The surface code or toric model can be alternatively represented as a doubled version of a $Z_2$-gauge theory. The gauge theory has the Hamiltonian $H=-\Delta \sum_p B_p$, and the commuting star operators $A_s$ are the local gauge symmetries of the Hamiltonian that flip the spins surrounding a vertex of the lattice. Wegner defined a 3D version of this model on a cubic lattice with 4-qubit plaquette terms $B_p^z, B_p^x,B_p^y$ orthogonal to the $z,x$ and $y$ axis respectively. He observed that this 3D model has a finite-temperature phase transition but no spontaneous symmetry breaking of a local order parameter such as the magnetization in the Ising model. In modern days this 3D Wegner model can be viewed as a three-dimensional version of the toric code. This 3D toric code was analyzed in detail in \cite{CC:3Dtoric}. The 3D toric code has an energy barrier scaling with lattice dimension $L$ for the logical $\overline{X}$ operator. What this means is that at sufficiently low temperature, below the critical temperature, a qubit initially encoded in $\ket{\overline{0}}$, will remain in the state $\ket{\overline{0}}$ modulo small fluctuations, i.e. small local $X$ errors. The candidate for the nonlocal order {\em operator} would then be $\overline{Z}$, whose expectation however depends on microscopic detail, the presence of small $X$ errors, and is therefore not expected to be stable. One can define an {\em error-corrected} order operator $\overline{Z}_{ec}$ \cite{AHHH:4d}  in which the process of correction against $X$ errors is taken into account, similar as the magnetization is robust against small sets of spin flips. Below the phase transition, the expectation of this operator $\overline{Z}_{ec}$ is expected to be stable \footnote{Most of the discussion on determining a nonlocal order parameter has been focused on finding a single parameter which distinguishes the low-temperature topological phase from the high-temperature phase. Characterizing the topological phase by a set of stable nonlocal order operators $(\overline{Z}_{ec},\overline{X}_{ec})$ whose expectations both vanish in the high-temperature phase, seems a proper quantum generalization of magnetization in classical memories.}.

The crucial departure from gauge theories comes from realizing that this is only half the story. We get the 3D toric code by adding the gauge symmetries to the 3D gauge theory so that $H=-\Delta (\sum_{p} \sum_{i=x,y,z} B_{p}^i+\sum_s A_s)$ where the star operators $A_s$ are now $X$-parity checks on the 6 qubits on edges emanating from a vertex in the 3D cubic lattice. In order to be thermally protected {\em against both} $\overline{X}$ and $\overline{Z}$ errors, we demand that in {\em both} gauge theories there is "spontaneous-symmetry breaking characterized by a non-local order parameter" below a critical temperature. This means that below the phase transition, both $\overline{X}_{ec}$ and $\overline{Z}_{ec}$ should have stable expectations. To understand this, imagine that we prepare the quantum memory initially in the state $\frac{1}{\sqrt{2}}(\ket{\overline{0}}+\ket{\overline{1}})$. Dephasing of this state would of course not affect the expectation of $\overline{Z}_{ec}$ (which for this state is $\langle \overline{Z}_{ec} \rangle=0$). But complete dephasing to the mixed state $\frac{1}{2}(\ket{\overline{0}}\bra{\overline{0}}+\ket{\overline{1}}\bra{\overline{1}})$ lets $\overline{X}_{ec}$ go from an initial value of $+1$ to $0$. 

Alas, in the 3D model, the star $A_s$ operators detect, similar as in the 2D surface code model, the end-points of strings of $Z$-errors. Such $Z$-error strings can grow and merge without energy expenditure leading to an $O(1)$ energy barrier for a logical $\overline{Z}$ error. In this sense the 3D toric code can be viewed as a model of robust storage of a classical bit which is protected against bit-flip errors $\overline{X}$ but not phase-flip errors $\overline{Z}$.
 
That there are dimensional obstructions to realizing thermally-stable topological order \cite{Nussinov:2008}, or finite-temperature robust storage of quantum information, or self-correcting quantum memories \cite{bacon} was also confirmed in \cite{BT:nogo}. These results prove the existence of an $O(1)$ energy barrier for any 2D stabilizer quantum code. A 4D version of the toric code model does exhibit all desired features and it has been shown that contact with a heat-bath would allow for a storage time $\tau$ growing exponentially with system-size below the critical temperature \cite{AHHH:4d}, a true demonstration of macroscopic quantum coherence, however in four spatial dimensions....

So, is this an answer to the question formulated in the article? Quantum information is intrinsically less robust than classical information as we live only in three spatial dimensions, whereas we need more dimensions quantumly as we need to be protected from both $X$ as well as $Z$ errors. Of course, such a simple picture is appealing but may be ultimately misleading. There are at least three caveats, from a pure theoretical perspective. 

First of all, we have a vision of active error correction for 2D quantum systems encoded in the surface code (or similar 2D topological quantum codes). This is the surface code architecture \cite{dennis+:top,RHG:threshold,fowler+:unisurf} in which one actively determines the presence and location of excitations. Ideally this architecture is implemented at the most physical level, that is, with a naturally suited dissipative mechanism which drives or keep the system in the code space without inducing logical errors.
One could say that we get around the dimensional obstruction by active inclusion of the time-dimension and classical processing. This machinery of error correction will be challenging to implement but several qubits under construction, such as ion-trap and superconducting transmon qubits, are coming close the noise rate of $10^{-4}$ which would be required to start benefitting from this scheme. 

Secondly, the situation in 3D concerning quantum self-correction, i.e. the presence of system-size dependent energy barriers for topologically-ordered many-body systems, is not clearcut. Results in \cite{BT:nogo} left open the question whether there would exist 3D lattice models where both logical operators were surface-like, even though this would seem hard to realize. On the other hand, from the picture of the toric code family, one can envision a certain no-go result, based on dualities. In $D$ dimensions, if one logical operator, say $\overline{X}$, has $d \leq D$-dimensional support, then $\overline{Z}$ is only left with at most $D-d$-dimensional support. This duality was proved for 3D translationally-invariant stabilizer codes in \cite{yoshida:dual}, under the restriction that the number of qubits encoded in the code is independent of lattice size (and thus only dependent on topology). Again, this leaves the possibility that there are 3D codes in which the number of encoded qubits depends nontrivially on lattice size and whose logical operators divide the available 3D space more equally. How? As Haah and later Haah and Bravyi showed in a 3D model by having logical operators with {\em fractal} support. The Haah model has an energy barrier proportional to $\log L$ \cite{BH:barrier}. From numerical and analytical studies, Bravyi and Haah found that the storage time $\tau \sim L^{\frac{c}{T}}$ for a constant $c$, thus increasing with system size $L$ as long as $L$ is less than some critical size $L^* \sim e^{\frac{3}{T}}$ \cite{BH:storage}.

Last but not least, the spin models that we consider are gross oversimplifications, or toy models, of the physical reality of interacting fermions and bosons and details of these more realistic models can crucially matter. For example, we have seen a different minimal dimension for spontaneous magnetization for the ferromagnetic Heisenberg model, namely three dimensions, and the Ising model, namely two dimensions. This difference relates to the fact that in the Heisenberg model, one breaks a local continuous symmetry (the commutation of the Heisenberg Hamiltonian with $U^{\otimes n}$ for any unitary $U$) while in the Ising model one breaks a discrete symmetry (the commutation of the Ising Hamiltonian $H$ with $X^{\otimes n}$) \footnote{In a quantum memory model encoding one qubit, the symmetry is continuous but non-local, i.e. $H$ commutes with any $\overline{U}=e^{i \theta \hat{n} \cdot \overline{S}}$ with $\overline{S}=(\overline{X},\overline{Y},\overline{Z})$ and the system is gapped unlike in models with local continuous symmetries.}. Another example is the dimensional dependence of superconductivity in which a non-zero value of the superconducting order parameter is only thermally stable in three dimensions. In two dimensions, a finite-temperature phase transition, a so-called Kosterlitz-Thouless transition, does still occur due to the attractive (logarithmically-scaling in distance) interaction between excitations (vortex anti-vortex pairs) \cite{book:tinkham}, which counteracts the entropic contribution below the transition temperature. 

Ultimately, the question of how to build a stable quantum memory will be decided in the lab, but I hope that our theoretical understanding can guide us in identifying approaches and physical systems which are likely to lead to success. I thank David DiVincenzo for interesting discussions clarifying some finer points of condensed matter physics.

\bibliographystyle{splncs}
\bibliography{fragbib}
\end{document}